\documentclass[aps,showpacs,twocolumn,nobibnotes]{revtex4-1}
\usepackage{graphics,graphicx,amsfonts,amsmath,amsbsy,amssymb,color}
\usepackage{bm}
\usepackage{subfigure}
\usepackage{multirow}
\usepackage{vector}  % Allows "\bvec{}" and "\buvec{}" for "blackboard" style bold vectors in maths

\def \beq {\begin{eqnarray}}
\def \eeq {\end{eqnarray}}
\def \Schrodinger {{Schr\"{o}dinger }}
\def \Di {{D_{\bfi}}}
\def \Dj {{D_{\bfj}}}

\def \bfj {{\bf j}}
\def \bfi {{\bf i}}

\def \mEh {{\textrm{mE}_{\textrm{h}}}}
\def \Eh {{\textrm{E}_{\textrm{h}}}}
\def \bohr {{\textrm{a}_{\textrm{0}}}}
\def \invbohr {{ \textrm{a}_{\textrm{0}}^{-1} }}
\def \tworonetwo {{ [2]_{\textrm{R12}} }}
\def \twos {{ [2]_{\textrm{S}} }}
\def \nadd {{n_{\textrm{add}}}}

\newcommand{\bra}{\ensuremath{\langle}}
\newcommand{\ket}{\ensuremath{\rangle}}
\def \ham {{\hat{H}}}

\newcommand{\rff}[1]{{Eq.~\eqref{#1}}}

\begin{document}
\title{An explicitly correlated approach to basis set incompleteness in Full Configuration Interaction Quantum Monte Carlo}
\author{George~H.~Booth$^{(a)}$}
\email{ghb24@cam.ac.uk}
\author{Deidre~Cleland$^{(a)}$}
\author{Ali~Alavi$^{(a)}$}  
\author{David~P.~Tew$^{(b)}$}
\affiliation{$^{(a)}$University of Cambridge, Chemistry Department, Lensfield Road, Cambridge CB2 1EW, United Kingdom}
\affiliation{$^{(b)}$School of Chemistry, University of Bristol, Bristol BS8 1TS, United Kingdom}

\begin{abstract}
By performing a stochastic dynamic in a space of Slater determinants, the 
Full Configuration Interaction Quantum Monte Carlo (FCIQMC) method has been able to obtain energies which are essentially
free from systematic error to the basis set correlation energy, within small and systematically improvable errorbars.
However, the weakly exponential scaling with basis size makes converging the energy with respect to basis set 
costly and in larger systems, impossible.
To ameliorate these basis set issues, here we use perturbation theory to couple the FCIQMC wave function to an explicitly correlated strongly orthogonal basis of geminals, 
following the $\tworonetwo$ approach of Valeev {\em et al.}. The required one- and two-particle density matrices are computed
on-the-fly during the FCIQMC dynamic, using a sampling procedure which incurs relatively little additional computation expense.
The F12 energy corrections are shown to converge rapidly as a function of sampling, both in imaginary time, and number of walkers. 
Our pilot calculations on the binding curve for carbon dimer, which exhibits strong correlation effects as well as substantial basis set
dependence, demonstrate that the accuracy of the FCIQMC-F12 method surpasses that of all previous FCIQMC calculations, and that 
the F12 correction improves accuracy equivalent to increasing the quality of the one-electron basis by two cardinal numbers.
%However, with exponential scaling with basis size, albeit weakly\cite{CBA2011}, it can be costly or impossible to converge the basis set 
%energy to the complete basis set limit, despite the highly accurate electron correlation treatment within the basis.
%To limit this need for brute force expansion of the basis, here we introduce a coupling to an explicitly correlated strongly orthogonal basis of geminals, following the $\tworonetwo$ approach
%of Valeev {\em et al.}\cite{Torheyden:JCP131-171103,Kong:JCP133-174126}. 
%These geminals couple perturbatively to the FCIQMC wavefunction via the two-electron density matrix, which is sampled and averaged on-the-fly during
%the FCIQMC dynamic, with relatively little additional computation. The convergence of the F12 basis set corrections are investigated as a function of sampling, 
%both in imaginary time, and number of walkers. Finally, the carbon dimer is investigated, surpassing the accuracy of previous FCIQMC calculations\cite{BoothC2}, in 
%a system that exhibits both strong correlation effects, as well as substantial basis set incompleteness.
\end{abstract}
\date{\today}
\maketitle

\section{Introduction}

The Full Configuration Interaction Quantum Monte Carlo (FCIQMC) method has 
arisen over the last few years as a way to obtain near exact ground state energies (in the Full Configuration Interaction sense)
for Hamiltonians in a given Hilbert space.
A range of Hamiltonians have been tackled from molecular\cite{BTA2009,CBA2010,BA2010,CBA2011,BoothC2} to solid-state systems\cite{Shepherd2012_1,Shepherd2012_2,Spencer2012}, obtaining
accurate correlation energies in Hilbert spaces far beyond the reach of traditional iterative diagonalizers.
%Systems spanning over 10$^{100}$ many-body
%basis functions have been considered, which are far beyond the reach of more traditional
%iterative diagonalizers, and even some other more sophisticated methods which attempt
%to approach this full configuration interaction (FCI) limit. 
This is achieved by stochastically sampling the underlying Slater determinant space using a discrete `walker' representation, 
thereby reducing the computational effort spent on the realisation of the large number of 
%This is achieved by
%discretizing the wavefunction components, and sampling them stochastically, thereby
%reducing the computational effort spent on the realization of the large number of 
low-weighted amplitudes in the expansion. The price of this stochastic algorithm is
the introduction of random errors, which nonetheless can be controlled and reduced as the
inverse square-root of the computational effort.

However, it has been known since the early days of electronic structure theory, that the necessity for these
huge many-body expansions is to a large extent an artifact of the basis
functions that are used to represent the wavefunction. The Full Configuration Interaction (FCI) expansion constructs the wavefunction
as a linear combination of all possible antisymmetrized spin-orbital products, called Slater determinants\cite{Handy1980,Knowles1984,Olsen1988}.
Although the evaluation of matrix elements, and the resulting linear optimisation problem is
relatively straightforward and computationally efficient, the size of the determinantal basis scales binomially with
the number of spin-orbitals, which is the source of the current restriction on FCI-based methodology.
Moreover, since the spin-orbitals are intrinsically
one-electron functions, the Slater determinant basis is ill-suited to some aspects of many-electron correlated motion, 
specifically those caused by the
singular electron-electron interactions at short range. This results in a slow convergence of the electronic 
energy with the size of the underlying basis, hence exacerbating the binomial bottleneck.

Due to this high scaling with basis size, enlargement of the basis set and subsequent extrapolation is 
in many cases impossible, particularly for the multi-configurational methods required for strongly correlated systems. 
% or simply: This high scaling with basis size, means that basis set enlargement and subsequent extrapolation 
%is in many cases impossible.
%
%This high scaling with basis size for multi-configurational methods which are required for
%strongly correlated systems, means that basis set enlargement and subsequent extrapolation 
%is in many cases impossible. 
As a consequence, for high-level methods, the basis set error is often far larger than the error 
in the correlation treatment within the basis. 
Even if convergence with respect to basis size is possible, strong correlation effects are generally considered to be a `small' basis set problem, and thus a similar
high-level treatment with increasingly large basis sets is an unnecessary burden.
Traditionally, this issue has been dealt with by partitioning the orbital space into a small, 
active basis in which the wavefunction is constructed from a multiconfigurational FCI-type expansion\cite{werner:mcscf,KnowlesMCSCF,Roos1980}, before the dynamical, short-ranged 
correlation is captured via excitations from this wavefunction into a larger, external basis. This is the rationale
behind methods such as complete active space second-order perturbation theory (CASPT2)\cite{RoosCASPT2,Andersson1992}, multireference configuration 
interaction (MRCI)\cite{Werner1988,WernerMRCI_2}, canonical transformation theory (CT)\cite{CT_Chan2006,CT_Chan2007} and others\cite{AngeliNEVPT2,KnowlesCIPT2}. 
However, these methods can be expensive, and problems can result if the strong correlation is not adequately captured by the active basis. 

In this paper, we take a
different approach, and attempt to overcome the slow basis set expansion by including geminal functions in the
expansion, which perturbatively couple to the FCIQMC wavefunction.
Hylleraas's pioneering work in 1929 demonstrated that a highly compact form for the 
short-ranged wavefunction behavior can be found by introducing an explicit dependence on the 
interelectronic coordinate, $r_{12}$\cite{Hylleraas1929,Hylleraas1928,Slater1928}. This naturally describes the dynamical correlation problem, which
covers both the exact form of the `cusp' at $r_{12}=0$, and the longer ranged coulomb hole around each electron
pair coalescence point. These cusp conditions, derived initially by Kato\cite{Kato:CPAM10-151,Pack:JCP45-556,Sahni2003}, result 
from the singularities in the coulombic electron-repulsion potential, which are balanced by
derivative discontinuities in the wavefunction. Neither these cusps, nor the short to medium-range coulomb holes are well approximated
by an expansion of the wavefunction in terms of Slater determinants formed from Gaussian-type orbitals, that are centered on
the nuclei of the system.

The umbrella of `explicitly correlated' methods covers ways to introduce this explicit dependence on $r_{12}$
into the wavefunction ansatze, including the use of exponentially correlated Gaussians\cite{Boys1960} and the transcorrelated method\cite{Handy69,Hattig:CR112-4}.  
These methods have fewer variational parameters to optimize, however, the presence of many-electron (more than two) 
integrals, and a more complicated optimization problem has restricted these methods to small systems, and their use is not yet routine.  
%Despite far fewer variational parameters to optimize, the presence of many-electron (more than two) 
%integrals, and a more complicated optimization problem has restricted these methods to small
%systems, and their use is not yet routine.  
An exception to this is the variational Monte Carlo method, where 
the need for integral evaluation is avoided altogether, by direct stochastic evaluation of the local energy
for the optimization of a non-linear Jastrow form for the wavefunction\cite{Foulkes2001}.

The R12/F12 methods\cite{Tew:BOOK2010,Werner:BOOK2010,Kong:CR112-75} augment the set of Slater determinants with a small number of determinants
that include two-electron geminal functions\cite{Kutzelnigg:TCA68-445,Kutzelnigg:JCP94-1985,Klopper87}. The resulting many-electron integrals 
are avoided, or approximated via resolution of the 
identity (RI)\cite{Kutzelnigg:TCA68-445,Kutzelnigg:JCP94-1985,Noga:JCP101-7738,Klopper:JCP116-6397,Valeev:CPL395-190}.
The result is a method, which for the same level of theory approaches the basis set limit much more rapidly\cite{Tew:PCCP9-1921}, 
thus greatly reducing the computational effort required for a given target accuracy.
This approach has been in active development for many years, and is now at a highly advanced stage
for single reference methods, where it is routinely used in a black-box fashion, and present in many quantum 
chemistry packages. However, apart from some early work\cite{Gdanitz:CPL210-253,Gdanitz:CPL283-253}, 
multi-reference F12 methods have only appeared relatively 
recently\cite{Ten-no:CPL446-175,Torheyden:JCP131-171103,Kong:JCP135-214105,Shiozaki:JCP131-141103,Shiozaki:JCP134-034113,Kedzuch:CPL511-418,Ondrej:PCCP14-4753,Haunschild:CPL531-247,Yanai:JCP136-084107}, 
even though the F12 approach offers great benefits in these cases, where the scaling 
with respect to basis size is often at its most prohibitive.

In this paper, we extend the stochastic FCIQMC method to include a perturbative coupling from the FCI determinant amplitudes, to a set of 
Gaussian-type geminal functions which explicitly correlate {\em all} pairs of orbitals in the FCIQMC basis (OBS).
This is in contrast to single-reference methods, where only the occupied orbitals of the reference determinant are correlated.
We follow the {\em a posteriori} $\tworonetwo$ approach of Torheyden and Valeev\cite{Torheyden:JCP131-171103}, where the geminal basis is used to 
compute a basis set incompleteness correction for the dynamic correlation energy using second-order perturbation theory.
The geminal contribution is not optimized, but chosen to satisfy the derivative discontinuity at the electronic cusps, 
and accurately represent the shape of the coulomb hole. Furthermore, we neglect the effect of relaxation of the FCIQMC amplitudes
and directly use the one- and two-particle reduced density matrices computed from the FCIQMC procedure.
Since the geminals are evaluated at a different level of theory to those FCIQMC amplitudes in the OBS, strict variationality with
respect to the FCI complete basis set (CBS) is lost.
Nevertheless, we will demonstrate that the perturbative level of theory for geminal amplitudes is sufficiently accurate and
that convergence to the FCI-CBS limit is much faster with our hybrid approach, where FCIQMC captures the strong correlation effects from 
the OBS, while the cheaper, polynomially-scaling, but still multireference F12 corrections are used for the 
remaining dynamic correlation. This `diagonalize-then-perturb' approach offers a balanced description
of the electronic correlation, while only requiring the small overhead from accumulating the density matrices on-the-fly. 

The geminal basis corrects for the incompleteness in the description of pair correlations, but there remains an incompleteness 
due to the finite basis for the one-electron description. To remedy this, we modify the $\twos$ method of Kong and Valeev\cite{Kong:JCP133-174126} 
designed to correct for basis set incompleteness in CASSCF wavefunctions, to provide a similar multireference {\em a posteriori} 
approach to one-particle incompleteness. For single reference methods, it has been demonstrated numerically
that the one- and two-particle incompleteness errors are largely decoupled\cite{Kohn:JCP132-024101,Kong:JCP135-214105}. Indeed, the
proposed corrections are independent and can be simply applied additively.
When both one- and two-electron corrections have been applied to the FCIQMC energy, we will denote the result FCIQMC-F12.

However, obtaining these corrections is not as straightforward as for many other deterministic methods. The strength of the FCIQMC 
approach lies in the fact that only a small fraction of the space is occupied at any one iteration, and therefore accurate 
energies and wavefunctions only emerge after appropriate time-averaging of the walker dynamic. 
However, explicit averaging of the FCIQMC wavefunction to obtain appropriate density matrices would negate many of the advantages of the method, requiring substantially
increased storage and computational costs.
%computational resorces which
%would necessarily scale at least linearly with the size of the FCI space.
Therefore, accumulation of accurate time-averaged
one- and two-particle reduced density matrices of the FCIQMC wavefunction required for the corrections
must be performed `on-the-fly', in a manner that neither 
becomes a substantial computational burden nor reduces the parallelism of the algorithm. In addition, the additional storage requirements
of such an algorithm should ideally remain modest, and of course, must not scale with the size of the FCI space.
%that scale with the size of the FCI space.

In this paper, we will demonstrate the potential of the FCIQMC-F12 method by first considering the the simple case of the dissociation of
the hydrogen molecule, to demonstrate some salient features of the approach, and the necessity for 
multireference F12 corrections in the presence of strong correlation effects. 
We will then discuss the quality of the on-the-fly stochastically sampled reduced density matrices, and their convergence both with 
elapsed imaginary time, and number of walkers in the space. Finally, a detailed study is performed on the carbon dimer, a strongly
correlated molecule, which is chosen since much work with FCIQMC has already been performed for this system (see Ref.~\onlinecite{BoothC2}). 
%We assess the variation of the energies with respect to the length-scale parameter, $\gamma$, of the geminal function, as well 
%as convergence with respect to auxiliary basis size. 
Accurate results for the carbon dimer are 
presented, and the outlook for the approach considered.

\section{Methodology}

\subsection{FCIQMC recap}

A brief overview of the FCIQMC method is given here, with more details provided in 
Ref.~\onlinecite{BTA2009,BA2010,CBA2010,CBA2011,BoothC2,Spencer2012}.  
%(check this is formatted o.k?).
Assuming a basis of $M$ orthogonal one-particle orbitals ($\{p,q,\dots\}$), an explicitly antisymmetric basis of $N$-electron Slater determinants can be formed, $\{D_{\bvec{i}},D_{\bvec{j}},\dots\}$, 
which exactly span the FCI-space. Exact diagonalization of the Hamiltonian in this basis provides the FCI energy (defining the 
basis-set correlation energy), and wavefunction components
on each of the determinantal basis functions ($\{C_{\bvec{i}},C_{\bvec{j}},\dots\}$), though its drawback is
that the dimension of this matrix scales factorially with both $M$ and $N$. 
In FCIQMC, the description of the determinantal amplitudes is coarse-grained through a discrete `walker' representation of the wavefunction. This allows for
a compression of the instantaneous representation of the wavefunction, and corresponding reductions in computational effort and storage. A set of
stochastically realised rules are then iteratively applied to each walker, and the energy averaged over these iterations, until the desired convergence
is reached.

The master equations governing the stochastic walker dynamic can be derived by starting with the imaginary-time \Schrodinger equation,
\begin{equation}
\frac{\partial \Psi}{\partial \tau}=-\ham \Psi . \label{eqn:TDSE}
\end{equation}
By performing an integration to large imaginary time, $\tau$, the ground-state wavefunction is projected out from the set of all stationary solutions,
%By performing a long-time integration to large imaginary time, $\tau$, the ground-state wavefunction is projected out from the set of all stationary solutions,
in keeping with the approach of all projector-based methods. A set of coupled differential equations is found,
\begin{equation}
-\frac{\textrm{d} C_{\bvec{i}}}{\textrm{d} \tau}= (H_{\bvec{ii}} - E_{\bvec{0}} - S) C_{\bvec{i}} + \sum_{\bvec{j}\ne \bvec{i}} H_{\bvec{ij}} C_{\bvec{j}}  ,   \label{eqn:ITSESoln2}
\end{equation}
where $C_{\bvec{i}}$ are the determinant coefficients, and
$S$ is introduced as an energy-offset `shift' parameter, which acts as a population control parameter, and 
provides an estimate of the correlation energy when it is allowed to vary to keep the walker population constant.

The coefficients in \rff{eqn:ITSESoln2} are discretised, and represented as the signed sum of walkers on the determinant. The second term in the equation
is simulated each iteration as a stochastically realised spawning criterion between connected determinants, while the first represents a diagonal 
death step. Annihilation events occur at the end of each iteration between oppositely signed walkers residing on the same determinant. However, brute force application of
this procedure will only converge onto the ground state assuming that a system-specific number of walkers is exceeded, since annihilation events need to be numerous enough
in order to overcome the `sign problem' present in the sampling of the space. If this is achieved, then at convergence, the signed number of walkers on any determinant will be
proportional to their FCI coefficient. 

The number of walkers required to achieve convergence to FCI accuracy within small error bars is dramatically accelerated by
invoking a systematically improvable approximation, termed initiator or $i$-FCIQMC, which will be used 
exclusively in this work. In this, the growth
of the occupied determinant space is controlled, such that previously unoccupied determinants can only 
become occupied if they are spawned onto from a determinant with a 
population exceeding a preset parameter $\nadd$. The rationale behind this is that determinants with 
weights larger than $\nadd$ are likely to have their sign established
correctly with respect to the rest of the instantaneous wavefunction, and therefore the uncontrolled 
propagation of noise from competing signed solutions of the problem should be limited.
Since the walker population on any determinant is constantly changing, so does the space which can successfully spawn 
onto the unoccupied determinants, and as such the approximation
cannot be written as a simple change to a static Hamiltonian matrix, but rather one which utilises efficient error cancellation 
within a time-averaged dynamic. The dynamic rigorously
converges to the original scheme as the walker number is increased, or $\nadd$ is reduced, which has 
been shown in several systems to provide an exponential saving over the original formulation.

The energy can be extracted from the dynamic via a non-variational projection onto a reference wavefunction, $D_{\textbf{0}}$,
\begin{equation}
E_{\textrm{proj}} = E_{\textbf{0}} + \sum_{\bvec{i}} \langle D_{\bvec{i}} | H | D_{\bvec{0}} \rangle \frac{ \langle C_{\bvec{i}} \rangle}{\langle C_{\textbf{0}} \rangle} , \label{eqn:ProjE}
\end{equation}
which in this work is chosen to be the instantaneously largest weighted single determinant in the space (which was the Hartree--Fock determinant for
geometries close to equilibrium). Alternatively, once the walker population has equilibrated, the value of the shift parameter, $S$, which renders the 
propagated FCIQMC wavefuncion L$^1$ norm-conserving can also be used as an estimate for the correlation energy, which when averaged over imaginary time is denoted $E_S$. 
Due to the independence of this estimate from a reference state, it is 
generally used for the highly multireference stretched geometries.
As we will see later, it is also possible to calculate another energy estimate from the trace of the reduced
Hamiltonian with the sampled two-electron density matrix,
but this in not generally the estimate which is used, due to its convergence properties which are discussed later.

\subsection{$\tworonetwo$ and $\twos$}

The $\tworonetwo$ and $\twos$ methods were developed by Valeev and co-workers as
perturbative basis set completeness corrections that can in principle be applied 
to any electronic state for which the one- and two-particle reduced density 
matrices are available\cite{Torheyden:JCP131-171103,Kong:JCP133-174126,Kong:JCP135-214105}.
Here we apply these corrections to the situation where the FCIQMC correlation energy has been determined in a 
computational basis of spin-orbitals $\{p,q,r,s\}$.
Consider the Mukherjee--Kutzelnigg normal-ordered Hamiltonian
\begin{align}
  \hat H = E^{0} + \hat F + \hat G \,,
\end{align}
where $E^0 = \bra 0 \vert \hat H \vert 0 \ket $ is the energy of the reference zeroth-order FCIQMC wave function $ \vert 0 \ket$.
The effective Fock operator $\hat F$, in a formally complete one-particle basis denoted by orbital indices $\{\kappa,\lambda$\}, is given by
\begin{align}
  \hat F =& f_\kappa^\lambda \tilde a^\kappa_\lambda \,,\\
 f_\kappa^\lambda =&  h_\kappa^\lambda + g_{\kappa p}^{\lambda q} \gamma_q^p \, \label{eq:fock}.
\end{align}
where $\gamma_q^p$ is the one-particle reduced density matrix which spans the one-particle basis of $ \vert 0 \ket$, 
$g_{\kappa p}^{\lambda q} = \bra \kappa p | \lambda q \ket$ are antisymmetric electron repulsion integrals, 
$\tilde a^\kappa_\lambda $ are the elementary normal-ordered operators in the Mukherjee--Kutzelnigg sense, and Einstein summation convention is assumed. 
%$\kappa,\lambda$ are orbitals
%of a formally complete one-particle basis and einstein summation convention is used for repeated indices. 
The zeroth-order Hamiltonian is chosen to be
\begin{align}
\hat H^0 = E^{0} + \hat P \hat F \hat P + (1-\hat P) \hat F (1-\hat P) + \hat P \hat G \hat P
\end{align}
where $\hat P$ projects onto the computational basis $\{p,q\}$, used for the FCIQMC calculation.

The $\twos$ energy correction is computed by second-order perturbation theory in a basis of 
states generated by singly exciting from the reference $\vert 0 \ket$ into complementary auxiliary (CA)
orbitals $\{a',b'\}$, orthogonal to the computational basis $\{p,q\}$.
\begin{align}
\vert 1 \ket =& t_{a'}^{p} \tilde a_p^{a'} \vert 0 \ket \, \\
E_{\twos} =& f_p^{a'} t^q_{a'} \gamma_q^p \,,
\end{align}
where the amplitudes $t^q_{a'}$ are calculated from the solution of
\begin{align}
A_{b'q}^{pa'} t_{a'}^q =& -\gamma^p_r f^r_{b'}  \, \\
A_{b'q}^{pa'} =& \delta_{b'}^{a'} f_q^p ( \lambda_{qs}^{pr} - \gamma_s^p \gamma_q^r ) + f_{b'}^{a'} \gamma_q^p  \,.
\end{align}
$\lambda_{pq}^{rs} = \gamma_{pq}^{rs} - \gamma_p^r \gamma_q^s + \gamma_p^s \gamma_q^r$ is the two-electron cumulant and 
$\gamma_p^q$ and $\gamma_{pq}^{rs}$ are the one- and two-particle density matrices. We also investigated
the performance of the zeroth order Dyall Hamiltonian\cite{Kong:JCP133-174126}, but
found that the convergence was less numerically stable.

The $\tworonetwo$ energy correction is computed by second-order perturbation theory in an internally contracted 
basis of Gaussian geminal-containing Slater determinants
\begin{align}
\vert 1 \ket =& \frac14 t^{xy}_{pq} \vert {\tilde \Gamma_{pq}^{xy}} \ket \,, \\
\vert {\tilde \Gamma_{pq}^{xy}} \ket =& \frac12 \hat O R_{\kappa\lambda}^{xy} \tilde a_{pq}^{\kappa\lambda} \vert 0 \ket \,.
\end{align}
The operator $\hat O$ ensures the strong orthogonality (one-electron orthogonality) condition of the geminal basis to the reference wavefunction, and the matrix elements 
\begin{align}
R_{\kappa\lambda}^{xy} =& \hat S_{xy} \bra \kappa \lambda \vert \hat Q_{12} f(r_{12}) \vert xy \ket
\end{align}
are the representation of the projected Gaussian geminals in second quantization. The projector $\hat Q_{12} = 1 - P_1P_2$ is used in order to simplify the working equations, and the rational generator
$\hat S_{xy} = (\frac38 + \frac18 \hat p_{xy})$, where $\hat p_{xy}$ interchanges the \emph{spatial} 
components of the spin-orbitals $x$ and $y$\cite{Ten-no:JCP121-117,Tew:MP108-315}, is used to impose the appropriate cusp conditions. The Gaussian geminals $f(r_{12}) \vert xy \ket$
are defined by the orbital space $\{v,w,x,y\}$, which is chosen to be the full computational basis $\{p,q,r,s\}$ and the
correlation factor $f(r_{12})$ is a linear combination of six Gaussians that closely fit
an exponential $-\gamma^{-1}\exp(-\gamma r_{12})$ \cite{Ten-no:CPL398-56}, according to Ref.~\cite{Tew:JCP123-074101}.
The $\tworonetwo$ energy correction is given by the Hylleraas functional
\begin{align}
E_{\tworonetwo} &=& \frac12 V_{pq}^{xy} t_{xy}^{pq} + \frac{1}{16} t^{vw}_{rs} (B_{vw}^{xy} \gamma_{pq}^{rs} - X_{vw}^{xy} \Phi_{pq}^{rs}) t_{xy}^{pq}
\end{align}
where the many-electron integrals are approximated using standard R12/F12 manipulations and 
through RI insertions of the basis $\{p',q',r'\}$, defined as the union of the computational basis $\{p,q,r,s\}$ and the CA basis $\{a',b'\}$. This results in
the working equations
\begin{align}
V_{pq}^{xy} =& \frac12 (  v_{rs}^{xy}
-  g_{rs}^{ta'} \gamma_t^u r_{ua'}^{xy}
- \frac12 g_{rs}^{tu} r_{tu}^{xy} ) \gamma_{pq}^{rs} \\
B_{vw}^{xy} =& - r_{vw}^{rq} f_r^p r_{pq}^{xy}
 - r_{vw}^{a'q} f_{a'}^p r_{pq}^{xy}
 - r_{vw}^{pq} f_{p}^{a'} r_{a'q}^{xy} \nonumber \\
& + r_{vw}^{pa'} \gamma_p^q f_{q}^{r} \gamma_r^s r_{sa'}^{xy}
 - r_{vw}^{pa'} \gamma_p^q f_{a'}^{b'} r_{qb'}^{xy} \nonumber \\
& - r_{vw}^{p'a'} f_{p'}^{p} \gamma_{p}^{q} r_{qa'}^{xy}
   - r_{vw}^{pa'} \gamma_{p}^{q} f_q^{p'} r_{p' a'}^{xy} \nonumber \\
& + \tau_{vw}^{xy} + \frac12 x_{v^\ast w}^{xy}
+ \frac12 x_{vw^\ast}^{xy}  + \frac12 x_{vw}^{x^\ast y}
+ \frac12 x_{vw}^{xy^\ast} \nonumber \\
& - r_{vw}^{p'q'} k_{p'}^{r'} r_{r'q'}^{xy} \\
X_{vw}^{xy} =& x_{vw}^{xy}
-  r_{vw}^{ta'} \gamma_t^u r_{ua'}^{xy}
- \frac12 r_{vw}^{tu} r_{tu}^{xy} \\
\Phi_{pq}^{rs} =& P(pq)P(rs)( \gamma_p^r \gamma_q^t f_t^u \gamma_u^s
+ \frac12 \gamma_t^s f_u^t \gamma_{pq}^{ru} \nonumber \\
& + \frac12 \gamma_p^t f_t^u \lambda_{uq}^{rs} - \gamma_p^r f_t^u \lambda_{uq}^{ts}) \,,
\end{align}
where $P(pq) O_{pq} = O_{pq} - O_{qp}$,
$\vert x^\ast \ket = (f_{p'}^x + k_{p'}^x) \vert p' \ket$ and
$f$ and $k$ are the Fock and exchange matrix elements (see Eq.~\ref{eq:fock}).
The geminal integrals are defined through
\begin{align}
r_{p'q'}^{xy}  =& \hat S_{xy} \bra p'q' \vert f(r_{12}) \vert xy \ket \\
v_{p'q'}^{xy}  =& \hat S_{xy} \bra p'q' \vert r_{12}^{-1} f(r_{12}) \vert xy \ket \\
x_{vw}^{xy}    =& \hat S_{vw} \hat S_{xy} \bra vw \vert f(r_{12})^2 \vert xy \ket \\
\tau_{vw}^{xy} =& \hat S_{vw} \hat S_{xy} \bra vw \vert (\nabla_1 f(r_{12}))^2 \vert xy \ket 
\end{align}
The geminal amplitudes, $t_{xy}^{pq}$ are not optimized, but selected to satisfy the coalescence conditions, which due to the presence of the rational generator, reduce to 
$t_{xy}^{pq}=\delta_{xp}\delta_{yq} - \delta_{xq}\delta_{yp}$.
The above equations are essentially those of Kong, Torheyden and Valeev, adapted for the present purpose.
The spin-adapted formalism presented in Ref.~\onlinecite{Kong:JCP135-214105} recasts the equations in terms of spin-free orbitals, although also introduces an approximation to remove the appearance of 4-RDMs in the formal
theory when calculating the $B$ intermediate. This modification is not expected to change results greatly\footnote{E.~Valeev: Personal communication}.
Our implementation reduces exactly to the single reference MP2-F12/2C$\ast$ F12 correction\cite{Kedzuch:IJQC105-929,Werner:JCP126-164102,Bachorz:JCC32-2492} 
if the diagonal Hartree--Fock density matrices are used. 
All integrals were computed using the {\sc Dalton} program\cite{dalton} and were combined with stochastically generated
one- and two-particle density matrices from FCIQMC in a stand-alone program.
%The approximations made in the derivation of these methods are weak ...

\section{Dissociation of Hydrogen}

In this section, we provide a simple example to illustrate the importance of an F12 correction that takes into account the multireference 
nature of the underlying wavefunction.
The hydrogen molecule provides a simple two-electron system which nonetheless encapsulates many of the problems in {\em ab initio} quantum
chemistry. At equilibrium and compressed bond lengths, dynamic correlation dominates, and both one- and two-electron basis set
incompleteness is generally significant. However, as the bond is stretched, the system becomes intrinsically multireference, as unphysical
ionic terms in the Hartree--Fock determinant need to be canceled by a similarly weighted excited determinant. At the dissociation
limit, the system reduces to two independent one-electron hydrogen atoms, with no local correlation between the two electrons. By size
extensivity, the energy of this system must be exactly -1$\Eh$. However, even if the static correlation is exactly captured, 
this will still only be achieved in the complete basis set (CBS) limit
since there remains an incompleteness in the one-electron description, which will simply be twice the error in the Hartree--Fock energy 
of a hydrogen atom in the same basis. This therefore provides a good system to consider both dynamic and static correlation 
effects, as well as separating the basis set incompleteness in both the one- and two-electron part of the wavefunction.

\begin{figure}[t]
\begin{center}
\includegraphics[scale=0.475]{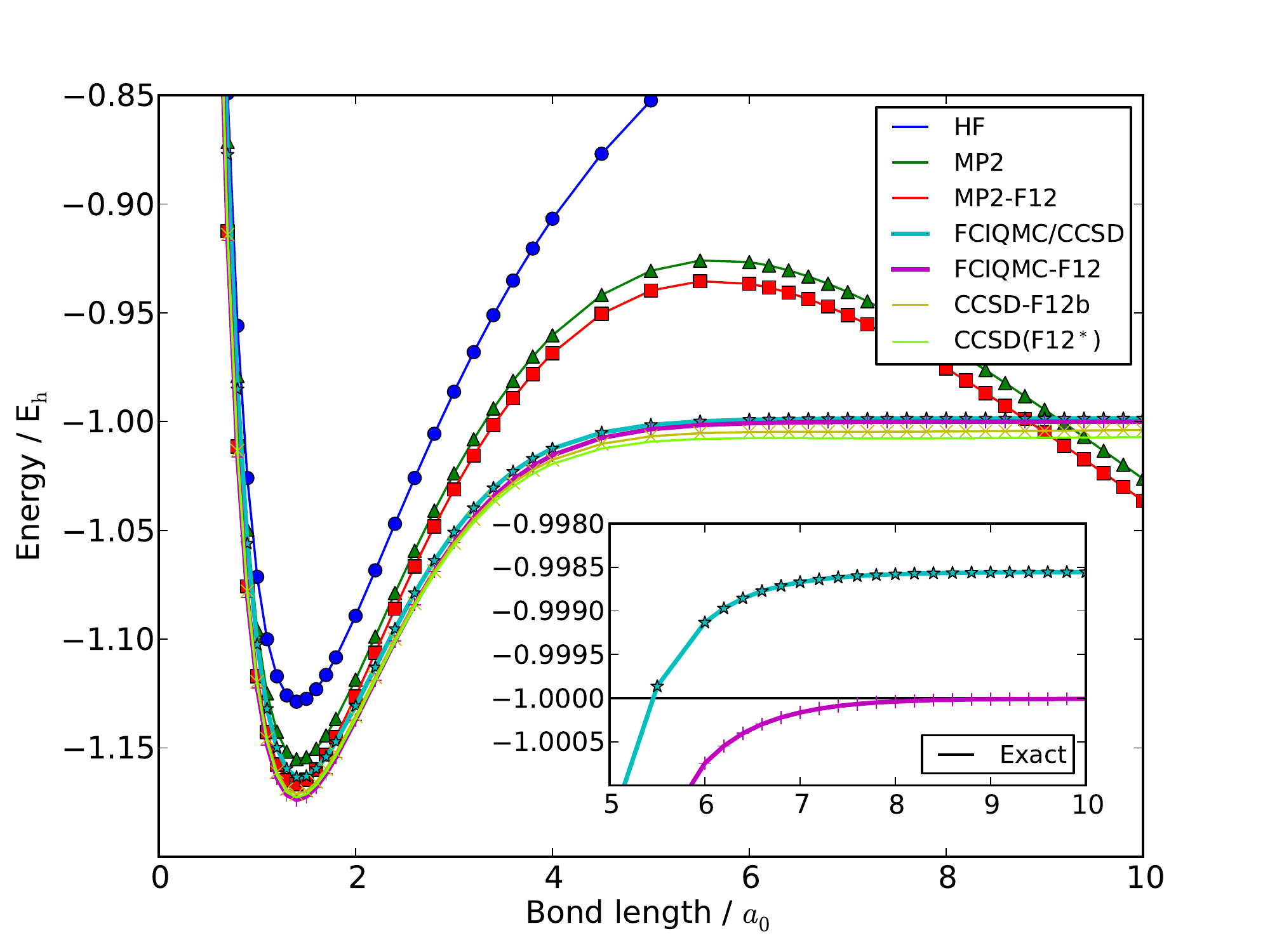}
\end{center}
\caption{Binding curves of H$_2$ in a cc-pVDZ basis. The system takes on a more multireference character as the bond is stretched.
Dissociation to two hydrogen atoms should yield an exact energy of -1$\Eh$. 
%MP2 gives a qualitatively incorrect description from bondlengths
%greater than $\sim2.5\bohr$. However, since this is a two electron system, CCSD agrees exactly with FCIQMC (and FCI) results 
%at all bondlengths. 
However, despite agreement between CCSD and FCIQMC, the single-reference F12 corrections all fail 
to successfully obtain the correct dissociation
limit, with only the fully multireference FCIQMC-F12 energy getting this 
exactly (error $\mathcal{O}[\mu\Eh]$). The individual corrections 
can be seen in Fig.~\ref{PlotH2Corrections}. All calculations used a cc-pV6Z-RI 
CABS basis, and a $\gamma$ of 1.0$\invbohr$. CCSD calculations
were run with {\tt MOLPRO}.
}
\label{PlotH2Binding}
\end{figure}

The binding curve for hydrogen in a cc-pVDZ basis is shown in Fig.~\ref{PlotH2Binding}. 
Second-order M\"oller--Plesset theory (MP2) gives a qualitatively incorrect description
for bond lengths greater than $\sim$$2.5\bohr$ due to the increasingly static correlation effects, whereas 
the Coupled-Cluster singles and doubles (CCSD) energies agrees exactly with FCIQMC (and FCI) values at all bond lengths, 
as expected for a two-electron system. However, the CCSD-F12 methods all fail to dissociate H$_2$ correctly,
with errors in the dissociation energy up to 7$\mEh$. The FCIQMC-F12 method achieves the exact dissociation 
energy to within $\mathcal{O}[\mu\Eh]$, which highlights the necessity for an F12 correction that explicitly correlates 
a multireference wavefunction.

\begin{figure}[t]
\begin{center}
\includegraphics[scale=0.475]{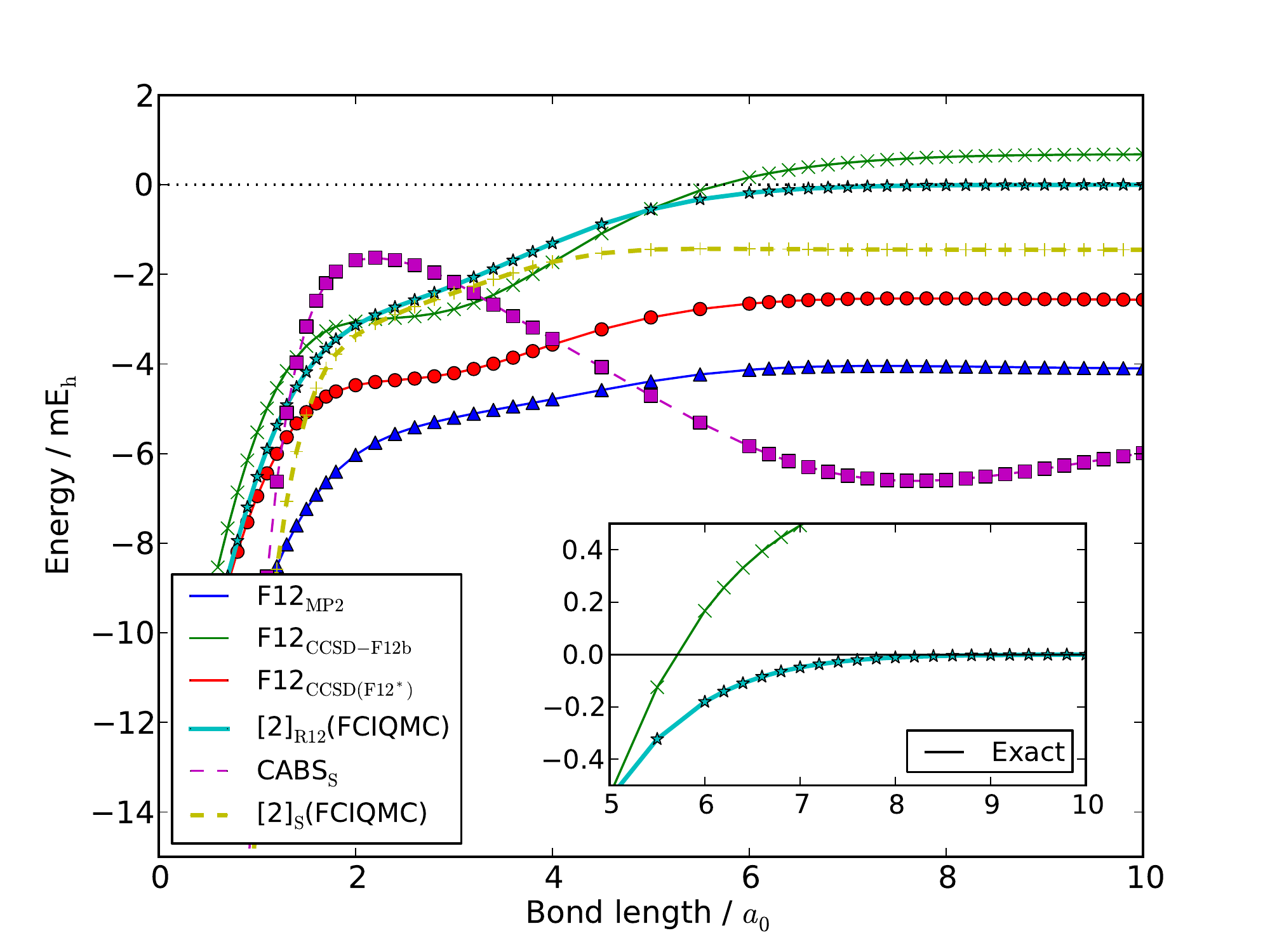}
\end{center}
\caption{Individual one- (dashed) and two- (solid) particle corrections to different methods across the dissociation of H$_2$, as given
in Fig.~\ref{PlotH2Binding}. Although there is no two-particle correlation at dissociation, only the $\tworonetwo$ F12 
correction to the FCIQMC
wavefunction correctly goes to zero in this limit. The multireference $\twos$ correction for single-particle incompleteness to the
FCIQMC wavefunction then provides the remaining energy to give the exact FCIQMC-F12 dissociation limit. The CABS singles
approach (CABS$_{\textrm{S}}$) overestimates the energy, while the single-reference F12 contributions to the 
different methods all converge to non-zero contributions.
}
\label{PlotH2Corrections}
\end{figure}

Further insight can be obtained from the individual contributions to the basis set incompleteness in Fig.~\ref{PlotH2Corrections}.
The reason for the failure of the CCSD-F12 methods is that the geminal amplitudes are selected such that they 
satisfy the coalescence conditions of the first-order wave function, that is, they explicitly correlate the 
Hartree--Fock orbital pairs only.
At infinite separation, the hydrogen molecule reverts to a sum of two atomic hydrogen atoms, and since the two electrons are now spatially separated
the two-electron F12 contribution should be rigorously zero. This spuriously does not occur for CCSD-F12, because
the Hartree--Fock determinant still contains terms with both electrons localized to the same atomic fragment.
CCSD-F12 variants where the geminal amplitudes are optimised rather than fixed do dissociate H$_2$ correctly, since the
geminal amplitudes optimise to zero at infinite separation, but these methods suffer from numerical instabilities 
and geminal basis set superposition errors\cite{Tew:JCP125-094302} and are not recommended.
In the FCIQMC-$\tworonetwo$ approach, all orbital pairs are explicitly correlated and the two electron correction 
naturally tends to exactly zero, as required. This example does not intend to highlight a deficiency of CCSD-F12 theory, since
for systems of more than two electrons the parent method would also fail for bond-breaking processes, but rather highlight the 
importance of an explicitly correlated geminal basis composed of more than the occupied Hartree--Fock orbitals in the presence of strong static correlations.

%\newpage

%===========================================
% H_2 binding VDZ 
%===========================================

%===========================================
% H_2 binding VDZ - corrections
%===========================================

Similar considerations apply to the one-electron basis set incompleteness correction, although this should not now go to zero at the dissociation limit. 
For increasing bond lengths the FCI basis set incompleteness tends to a sum of the Hartree--Fock basis set errors for the individual hydrogen atoms.
Again, the correction for this based on a single Hartree--Fock reference\cite{Knizia2007} is shown to be
inadequate for long bond lengths where strong correlation effects are large. 
However, the FCIQMC-$\twos$ approach almost exactly corrects for the one-electron error over the whole binding curve, 
resulting in a total energy at dissociation of -1$\Eh$.

\section{Reduced Density Matrices from FCIQMC}

In order to construct the perturbative coupling to the geminal basis, it is necessary to obtain the one- and two- electron reduced density matrices (RDM).
Much of the benefit of the FCIQMC approach would be lost if the time-averaged wavefunction over the entire FCI space was required in order to construct these, 
and even an explicit consideration of all occupied determinants connected via the desired RDM each iteration would become impractical when extending the 
approach to larger systems. Despite this, an accurate extraction of the RDM information from the time-averaged FCIQMC dynamics is essential for this
method, while many other important molecular properties, such as nuclear gradients, dipole moments and polarizabilities can be obtained from the density matrices
via a trace of the observable operator with the appropriate density matrix. 
%Many of these properties, along with details of the algorithm used in the construction
%of these matrices will be the subject of a forthcoming paper, while here we will be primarily concerned with the quality and convergence of the sampled RDMs in this 
%F12 approach, and the accuracy of the resultant energies.

The reduced density matrices over all orbitals in the OBS can be defined in second quantization as
\begin{align}
\gamma_p^q =& \langle \Psi | \hat{a}_p^{\dagger} \hat{a}_q | \Psi \rangle   \\
\gamma_{pq}^{rs} =& \langle \Psi | \hat{a}_p^{\dagger} \hat{a}_q^{\dagger} \hat{a}_s \hat{a}_r | \Psi \rangle ,
\end{align}
where the distinction between the one- and two-body matrices is evident from the rank of the tensor. Since the Hamiltonian operator that is sampled 
during the FCIQMC dynamic contains the same excitation rank as these matrices, it is possible to devise a way to sample the one- and two-body density
matrices on-the-fly and stochastically, during the FCIQMC dynamic.  
The contributions must the be appropriately unbiased for the magnitude of the Hamiltonian matrix element between individual connected determinant pairs.  
The precise manner in which this is done will be described elsewhere.
%as long as the contributions are appropriately unbiased for the magnitude of the Hamiltonian
%matrix element between individual connected determinant pairs. During the spawning step each iteration, the determinant population is encoded and packaged with newly spawned walkers to 
%the stochastically generated determinant connected via one and two electron excitations in each successful spawning attempt. 
%Here, the weight is combined with the population on the newly occupied determinant and summed into the accumulated density matrix.
%This process starts after an appropriate equilibration time, and continues for a number of iterations until sufficient statistical accuracy is obtained, before the matrices
The accumulation of the reduced density matrices begins once the simulation has reached equilibrium and continues for a number of iterations until 
sufficient statistical accuracy is obtained, before the matrices 
are normalized according to their required trace relation. As will be shown, in many cases the density matrices converge rapidly with iterations, 
resulting in a relatively small computational overhead for their accumulation.

Ignoring potential errors in the description of the wavefunction due to the initiator approximation, two errors can arise due to this on-the-fly sampling of the
density matrices. Since they are accumulated via a sampling of the Hamiltonian, it requires that a non-zero Hamiltonian matrix element
exists between all pairs of occupied determinants which are connected via one- and two- body excitation operators. If this is not the case, then no walkers can 
spawn between these determinants, and therefore no contribution can be made to the density matrices from this pair, introducing a 
bias into the matrices (but not the variational energy from the matrices). Determinants
between different symmetry blocks of the Hamiltonian are not connected, however since there will be no weight on determinants outside the currently sampled symmetry block,
this does not introduce an error. 

The only instance of this criteria being rigorously unfulfilled is in the case of Brillouin's theorem when the one particle basis
consists of canonical Hartree--Fock orbitals, and Hamiltonian matrix elements between the Hartree--Fock determinant and all of its single excitations are rigorously zero, despite
potentially significant weights on these determinants. To account for this, and improve the description of the RDMs, all connections between the Hartree--Fock determinant and its single
and double excitations, are considered explicitly during each iteration. This removes any error due to Brillouin's theorem, and improves the quality of the RDMs 
%from important contributions from the Hartree--Fock, 
with negligible computational overhead. The small possibility that two significantly weighted symmetry-similar determinants are connected via a single or double excitation operator,
but with a statistically zero Hamiltonian matrix element between them, remains a potential source of error.

The other possible error from this sampling of can arise if instantaneous wavefunction values are used, rather than walker populations averaged in time,
i.e. the assumption that $\langle C_{\bvec{i}} C_{\bvec{j}} \rangle = \langle C_{\bvec{i}} \rangle \langle C_{\bvec{j}} \rangle$, where the averages are 
over the iterations of the FCIQMC run during accumulation of the density matrices. This assumes no serial correlation between the 
errors in the determinant populations on $\Di$ and $\Dj$, and therefore
enough walkers such that any correlated fluctuations in the amplitudes do not introduce a bias. 
Since the RDMs are functions of products of these determinant coefficients, this error will not decay to zero simply with increasing sampling time.

However, as opposed to the first error,
this bias is systematically improvable, and will rigorously vanish in the limit of a large number of walkers. To ameliorate this error when not at this limit, an averaged walker population over the duration
of non-zero occupancy is maintained for each occupied determinant. Since this is only calculated over the instantaneously occupied subspace, it again involves only small additional effort, and
weights were not maintained across the whole space. However, an error is still present since this averaged walker population information is lost once a determinant becomes unoccupied. 
An important question is the rate at which any error decays as the total walker population increases, especially in comparison to the decay of the initiator error
and the size of the random errors. This rate of convergence is likely to depend on the specific property that the density matrices are being used to calculate, and in
this paper, we will consider the convergence of the error in the calculation of the F12 corrections and energy estimators with increased walker number, as well as the convergence
with increased iterations.

\subsection{Convergence of Density Matrix properties}

\begin{figure}
\begin{center}
\includegraphics[scale=0.45]{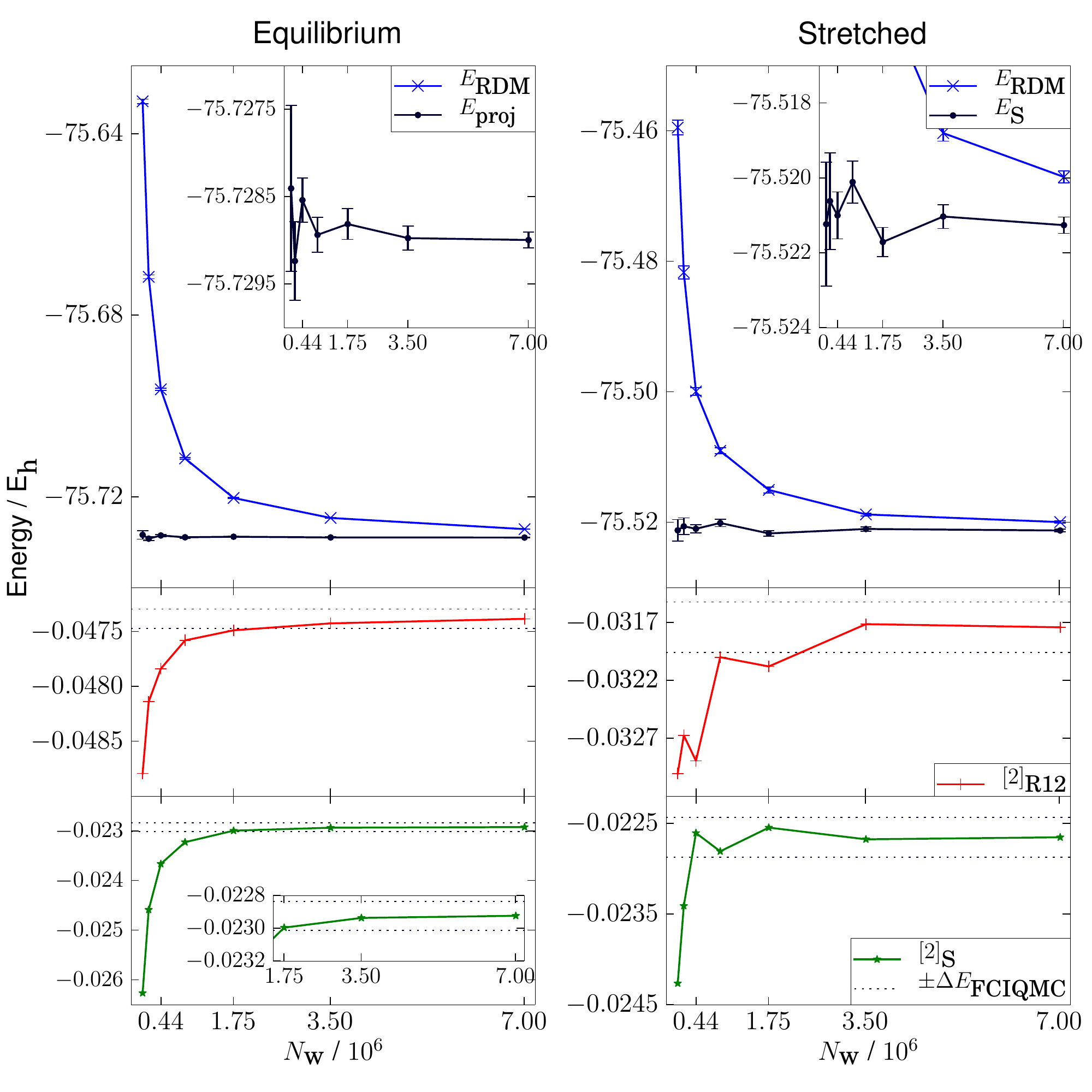}
\end{center}
\caption
{Convergence of the energy estimates and F12 corrections with increasing walker number, for C$_2$ in a cc-pVDZ basis set, 
at equilibrium (1.3~\AA) and stretched (7.0~\AA) geometries.  
The dotted lines represent the stochastic errors on the most accurate 
$E_{\textrm{proj}}$ or $E_{\textrm{S}}$ 
values, relative the final $\tworonetwo$ and $\twos$ results.
The density matrices were accumulated over 100,000 iterations.
%$N_{\textrm{FCI}} = 1.398 \times 10^7$ spin-coupled functions (no Lz).
} 
\label{Plot_C2_VDZ_Nw}
\end{figure}

\begin{figure}
\begin{center}
\includegraphics[scale=0.45]{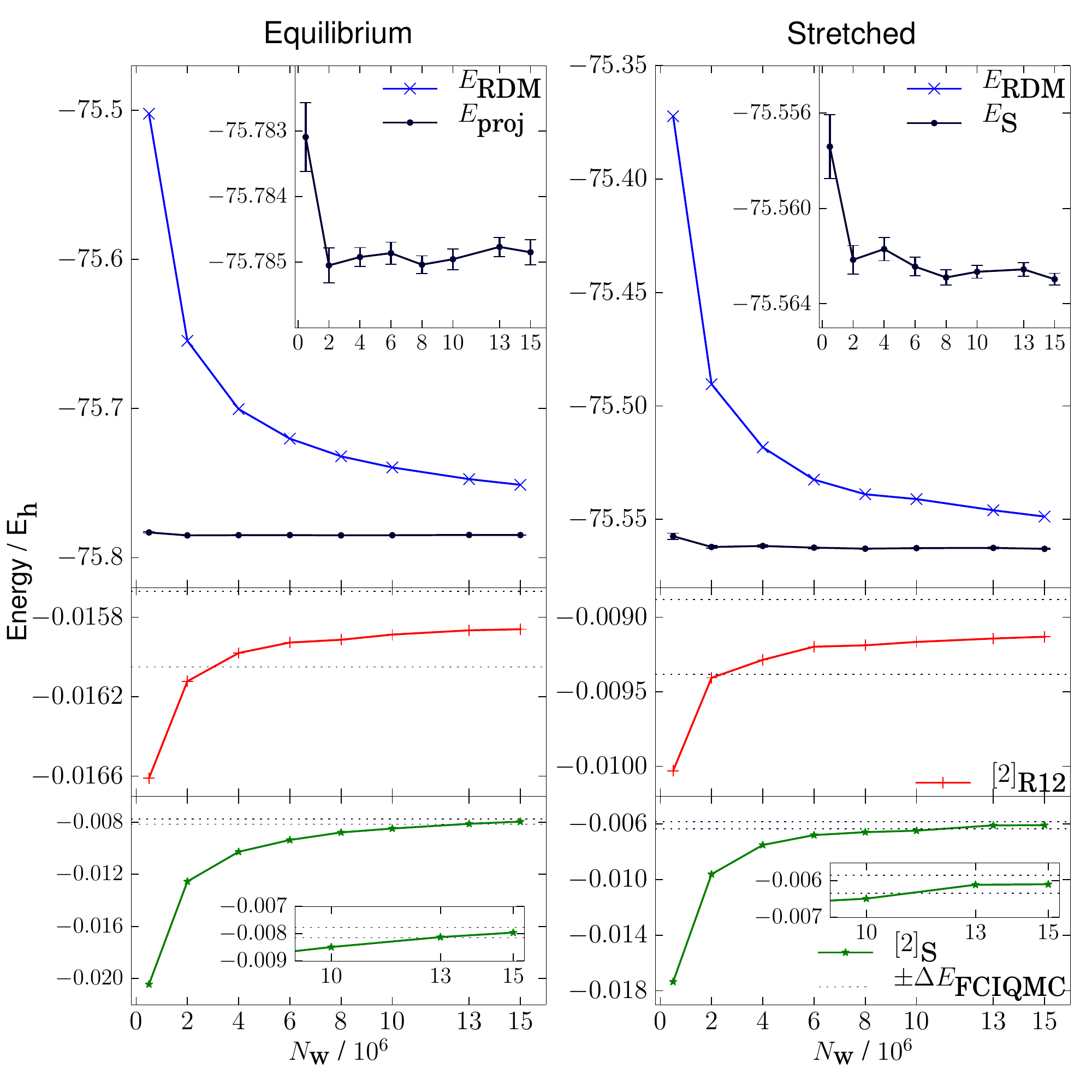}
\end{center}
\caption
{Convergence of the energy estimates and F12 corrections with increasing walker number, for C$_2$ in a cc-pVTZ basis, 
at equilibrium (1.2425~\AA) and stretched (7.0~\AA) geometries.  
%The projected energy and shift values appear to be converged at approximately $8 \times 10^6$ walkers.  
The dotted lines represent the stochastic errors on the most accurate 
$E_{\textrm{proj}}$ or $E_{\textrm{S}}$ 
values, relative the final $[2]_{\textrm{R12}}$ and $[2]_{\textrm{S}}$ results.
The density matrices were accumulated over 100,000 iterations.
%$N_{\textrm{FCI}} = 1.125 \times 10^{10}$ spin-coupled functions (no Lz).
}
\label{Plot_C2_VTZ_Nw}
\end{figure}

To investigate the convergence of the density matrices with walker number and imaginary time, we consider both the equilibrium (1.3~\AA) 
and stretched (7.0~\AA) geometries of the
frozen-core carbon dimer molecule in a restricted Hartree--Fock basis, with a geminal exponent fixed to equal 1.0$\invbohr$. 
These two bond lengths highlight vastly different correlation effects and wavefunctions,
to provide contrasting tests for the accuracy of the sampled density matrices. The equilibrium geometry is predominantly single-reference, 
dominated by the weight at the Hartree--Fock determinant, while the stretched case is highly multiconfigurational, with no single determinant 
contributing overwhelmingly to the electronic structure. Fig.~\ref{Plot_C2_VDZ_Nw} shows the convergence of various estimators used in this work 
with increasing walkers in a cc-pVDZ basis set, while Fig.~\ref{Plot_C2_VTZ_Nw} extends this to the larger cc-pVTZ basis. It should be noted that 
these calculations are independent to the ones calculated in Ref.\onlinecite{BoothC2}, since angular momentum symmetry was not explicitly included in these 
calculations. This resulted in a larger space of $2.25 \times 10^{10}$ symmetry-allowed determinants in the cc-pVTZ basis, although 
through the use of time-reversal symmetry, the number of distinct $N$-electron functions is about half this number.

The $E_{\mathrm{RDM}}$ energy estimator is calculated from the trace of the two-electron RDM with the reduced two-electron Hamiltonian,
\begin{equation}
E_{\mathrm{RDM}} = \sum_{pqrs} \gamma_{pq}^{rs} k_{pqrs} = \mbox{Tr}[\gamma k] , \label{eqn:RDMEnergy}
\end{equation}
where
\begin{equation}
k_{pq}^{rs} = \frac{1}{N-1} h_p^r \delta_q^s + g_{pq}^{rs} .
\end{equation}
This should not be confused with the projected or shift energy estimator from the FCIQMC calculation, which is used to calculate the energy in 
FCIQMC, and is defined in \rff{eqn:ProjE}. Generally, \rff{eqn:RDMEnergy} would define a strictly variational energy for the wavefunction, and be 
equivalent to the pure estimate $\langle \Psi | H | \Psi \rangle$. However, since the RDM is not explicitly calculated, but rather stochastically 
sampled, it may not strictly correspond to the FCIQMC wavefunction at all times. This therefore has the potential to break strict N-representability, 
and thus return a non-variational energy, however this is not found in this study and always represents a variational 
estimate. The effect of the correlated sampling error, compounded by the variational constraint, means that the $E_{\mathrm{RDM}}$ energy estimator is 
found to converge to the FCI energy of the system 
far slower with respect to walker number compared to the projected or shift energy, as shown in Fig. \ref{Plot_C2_VDZ_Nw}. For this reason, it is not
used as an estimate of the FCI basis set energy of the system.

%===========================================
% VDZ CONVERGENCE WITH WALKER NUMBER
%===========================================

However, this is not found to be the case for the perturbative basis set incompleteness corrections, with $\tworonetwo$ and $\twos$ based on the sampled density matrices also shown. These 
indicate convergence to approximately within the stochastic errors of the final FCIQMC energy estimate when the walker number exceeds $3.5 \times 10^6$, for 
both the equilibrium and stretched geometries. These errors were 0.17$\mEh$ in the projected energy ($E_{\textrm{proj}}$) for the 
equilibrium case and 0.43$\mEh$ in the shift estimate ($E_{\textrm{S}}$) for the stretched case. Intriguingly, 
the convergence is faster for the multiconfigurational stretched case, although less monotonic. In the stretched case, the two-particle F12 
correction is only $\sim$66\% of the correction at equilibrium, reflecting the increased separation of the electrons localized to each atom and thus 
reduced energetic importance of the cusps between them. However, the magnitude of the incompleteness in the one-electron
space is virtually unchanged, and is approximately half the size of the $\tworonetwo$ correction.

%===========================================
% VTZ CONVERGENCE WITH WALKER NUMBER
%===========================================

The convergence in the cc-pVTZ basis (Fig.~\ref{Plot_C2_VTZ_Nw}) shows a similar trend, with large errors remaining in the energy from the density 
matrices ($E_{\mathrm{RDM}}$), while the F12 corrections converge at a much faster rate. Convergence to approximately within the errorbars of the FCIQMC energy (0.19$\mEh$ for the 
equilibrium case and 0.25$\mEh$ for stretched) is achieved within a sampling of 15 million walkers. The magnitude of the $\tworonetwo$ basis set 
corrections are only approximately a third of the size in the cc-pVTZ basis compared to the cc-pVDZ basis, while the $\twos$ correction is reduced 
even further to approximately a quarter of its previous size. This reflects the increased correlation captured by FCIQMC in the larger orbital basis, 
while indicating that this improves the accuracy of the one-body description more than the two-body case.

The observation that the F12 corrections converge much faster than the energy estimate from the density matrices can be rationalized by
considering the limiting case of a diagonal Hartree--Fock density matrix. This would recover the single-reference F12 and CABS singles 
corrections of MP2, which are already quite accurate at equilibrium, whereas the correlation energy computed using $E_{\mathrm{RDM}}$ is
by definition zero.
It is key to note that in both cases, the convergence of the F12 corrections with walker number to within the stochastic errors of the $i$-FCIQMC 
energy is not significantly slower than the number of walkers required to acceptably remove the effects of the initiator approximation in the FCIQMC energy 
estimate, and therefore the additional expense of calculating the F12 corrections is not a large fraction of the overall computational cost. 

\begin{figure}
\begin{center}
\includegraphics[scale=0.45]{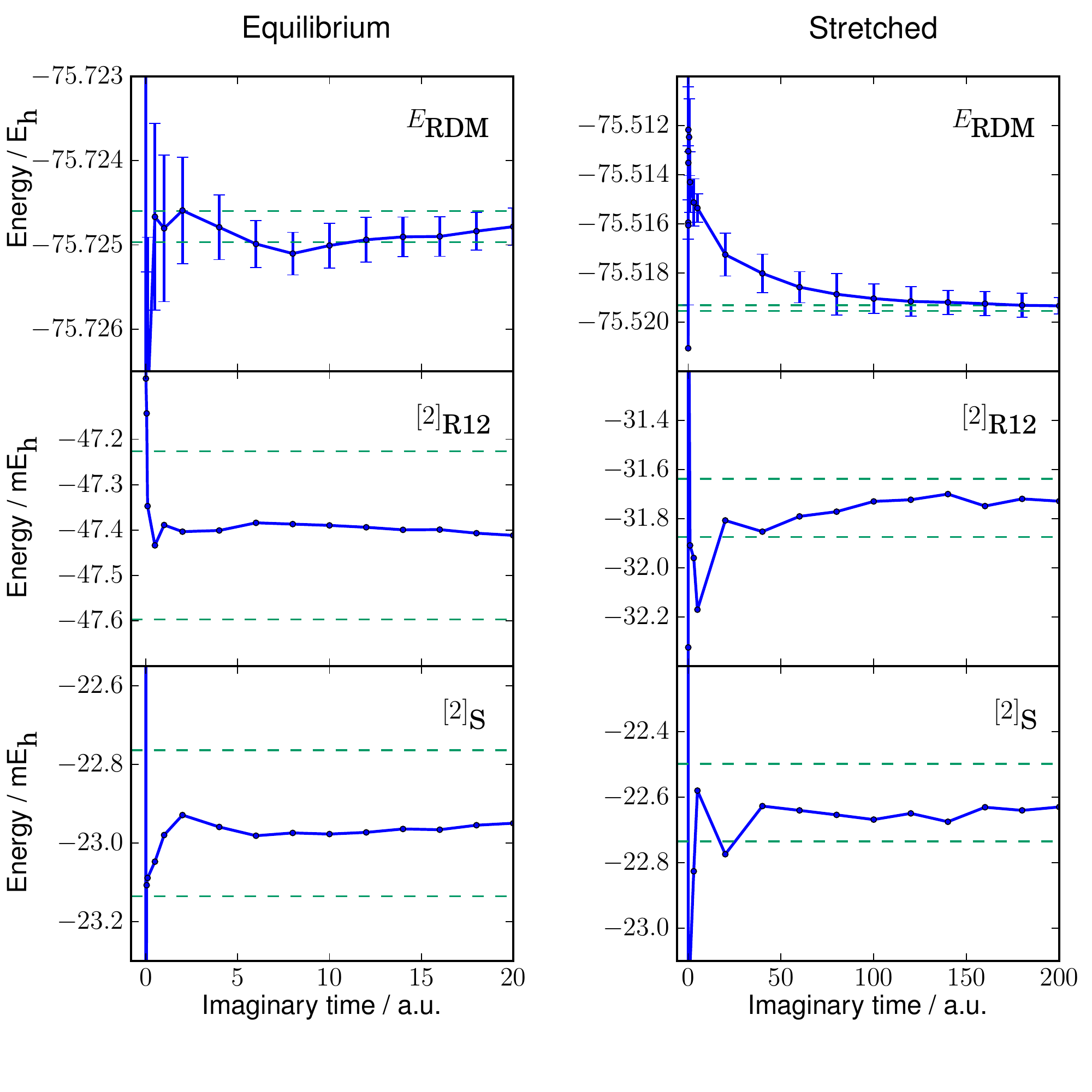}
\end{center}
\caption
{
Convergence of the energy estimates with imaginary time spent sampling the density matrices for C$_2$ in a cc-pVDZ basis for both equilibrium (1.3~\AA) 
and stretched (7.0~\AA) geometries. 3.5 million walkers sampled the space.
Dotted lines show typical error bars in FCIQMC projected energy and shift, taken to be the same as in Fig.~\ref{Plot_C2_VDZ_Nw}.
%Perhaps change this so the dotted lines change corresponding to errors calculated over the same imaginary time?
} 
\label{Plot_C2_VDZ_Iter}
\end{figure}

%\subsection{Convergence of Density Matrix properties with sampled imaginary time}
Another key consideration is the convergence of the F12 corrections with respect to sampled imaginary time.
Fig.~\ref{Plot_C2_VDZ_Iter} shows this for the cc-pVDZ basis, again at both equilibrium and stretched geometries.
In contrast to the convergence with walker number, significant differences exist between the two geometries, with the 
more multiconfigurational wavefunction at stretched geometry requiring many more iterations to reach energy convergence to within typical stochastic
errorbars of the $i$-FCIQMC energy estimator. Despite this, the number of iterations required to sample the density matrices to 
obtain this accuracy in the F12 estimates is again not generally more than than required to obtain typical errorbars in the calculation of the 
FCIQMC energy estimate ($\mathcal{O}[10^{-4}]\Eh$). Although not shown, the cc-pVTZ basis exhibits a similar convergence.

%===========================================
% VDZ CONVERGENCE WITH IMAGINARY TIME
%===========================================

\section{The Carbon Dimer}

The carbon dimer has been the focus of a recent FCIQMC investigation, where its strong correlation effects, state crossings of the same Abelian symmetry group, avoided crossings and
large basis set incompleteness provided a stringent test of any method\cite{BoothC2}. Despite enlargement of the basis to quadruple-zeta quality, it was evident that large basis set
errors remained. This is now tackled within the framework of the explicitly correlated approach in order to improve upon these results.
%Following the investigations into the required sampling of the density matrices required to converge the F12 corrections 
Following the convergence investigations shown 
in Figs.~\ref{Plot_C2_VDZ_Nw} and \ref{Plot_C2_VTZ_Nw}, the cc-pVDZ and cc-pVTZ C$_2$ binding curves were calculated with FCIQMC-F12 
using $3.5 \times 10^6$ and $13 \times 10^6$ walkers respectively, for each geometry.  The 
2-RDM was stochastically constructed once each simulation reached constant walker mode.  
%The cc-pVDZ and cc-pVTZ binding curves given by $E_{\textrm{proj}}$ and 
%$E_{\textrm{S}}$ are presented in Fig.~\ref{Plot_C2_F12_BindingCurves}.  
%
%From comparison of the cc-pVDZ energies to those presented in Ref. \onlinecite{BoothC2}, 
%these new cc-pVDZ results appear to relax onto the lower energy $B^1\Delta_g$ state for all 
%geometries between 1.7 and 2.7~\AA\ (inclusive), and thus describe the adiabatic potential energy surface. 
%This is important to ascertain, since in these calculations,
%angular momentum symmetry was not used, and so there was no way to exclude the $X^1\Sigma^+_g$ or $B^1\Delta_g$ from the space since they span the same irreducible
%representation of the D$_{\textrm{2h}}$ abelian point group.
%On the other hand, the cc-pVTZ results in Fig. \ref{Plot_C2_F12_BindingCurves} all converge to the $X^1\Sigma^+_g$ state
%(see Ref.~\onlinecite{BoothC2}), for all points except 2.2~\AA, despite the presence of the lower energy 
%symmetry allowed $B^1\Delta_g$ state at some geometries. 
%For the geometries for which the $X^1\Sigma_g^+$ state was obtained, the energies agree with those presented in 
%Ref.~\onlinecite{BoothC2} to within stochastic error bars.
%the $i$-FCIQMC simulations still converged onto the $X^1\Sigma_g^+$ solution.  This is similar to the effect 
%depicted in Figure \ref{Plot_C2_F12_BindingCurves}, 
%for the 6-31G$^{*}$ basis set.  
Angular momentum symmetry is not explicitly imposed in our calculations. Therefore,
to avoid complications arising from state crossings and metastable convergence to excited states\cite{BoothC2}, we considered bond lengths
less than 1.6~\AA\, where the ground $X^1\Sigma_g^+$ state is dominant, as well as geometries near dissociation with bond lengths more than five times
the equilibrium.

%%Since angular momentum symmetry was not conserved in the more recent calculations, to avoid complications regarding state crossings and metastable convergence to excited states\cite{BoothC2},
%%only geometries less than 1.6~\AA\ were considered, where the ground $X^1\Sigma_g^+$ is dominant, as well as the highly dissociated geometry of five times equilibrium. The 
%Since angular momentum symmetry was not conserved in the more recent calculations, 
%only geometries less than 1.6~\AA\ were considered, 
%to avoid complications regarding state crossings and metastable convergence to excited states\cite{BoothC2}.  
%At these bondlengths, the ground $X^1\Sigma_g^+$ is dominant.  The highly dissociated geometry of five times equilibrium was also included.  
%The non-parallelity errors (NPE), defined as the difference between the largest and smallest deviations within these geometries, can be seen in Fig.~\ref{Plot_C2_NPE_VTZF12} and Table~\ref{Tab:NPE}, 
%relative to the cc-pVTZ FCIQMC-F12 results. 
% Weird formatting going on around '(NPE)' and the dashes.

\begin{figure}
\begin{center}
\includegraphics[scale=0.45]{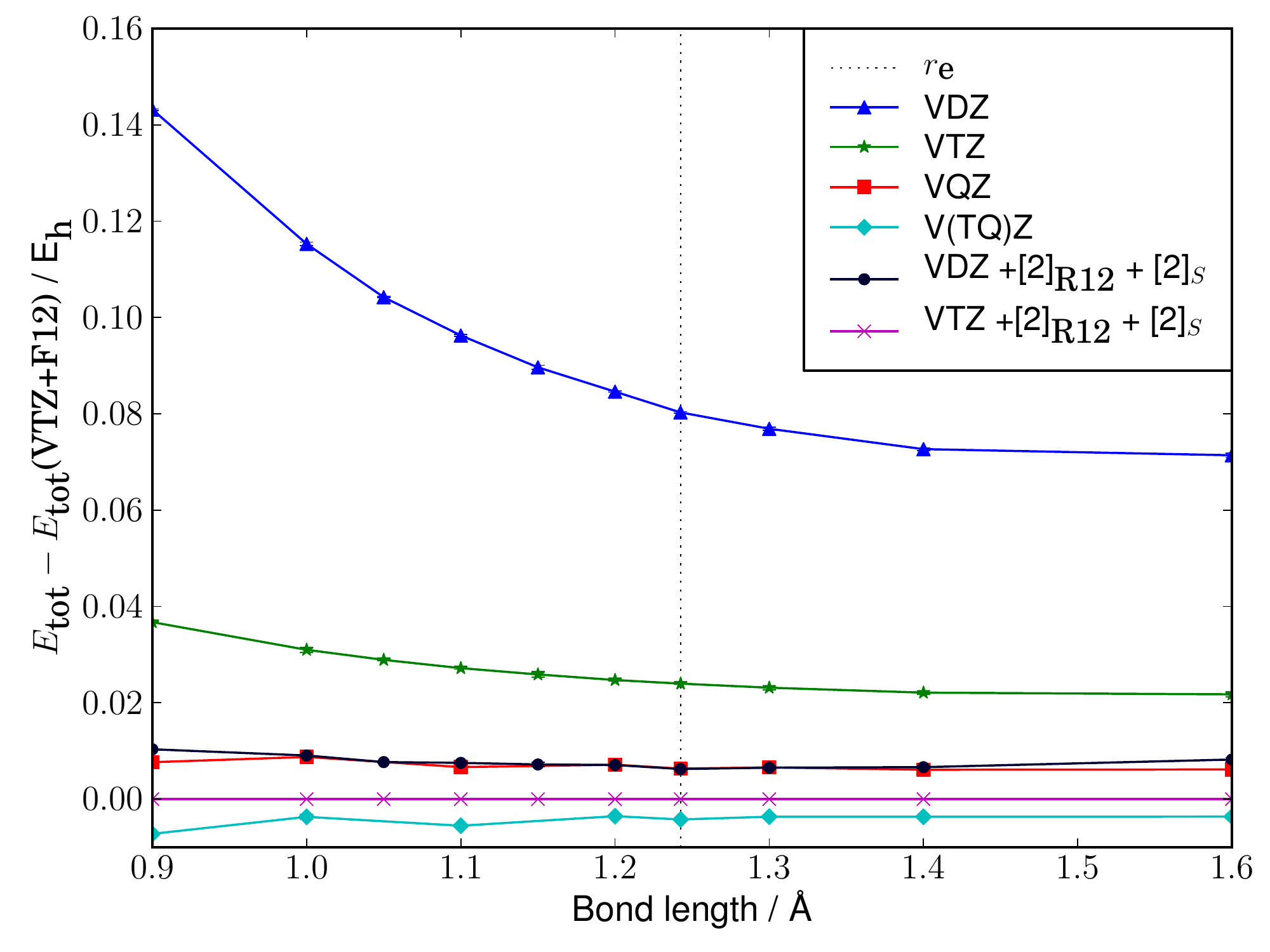}
\end{center}
\caption[]
{The errors relative to the FCIQMC-F12 results in the cc-pVTZ basis, for the C$_2$ binding curves calculated using the cc-pVDZ, cc-pVTZ and 
cc-pVQZ basis sets, as well as the cc-pVDZ energies with F12 corrections and extrapolated cc-pV(TQ)Z energies\cite{Halkier1998_Extrap}.  Only the 0.9 - 1.6 \AA\ region is shown, 
as all these calculations converge onto the $X^1\Sigma_g^{+}$ state. 
%This range contains the points with the biggest non-parallelity errors.  
With F12 corrections, the NPE across the entire cc-pVDZ binding curve is reduced from 86.8(4) to 12.3$\mEh$.  
The NPEs are compared in Table~\ref{Tab:NPE}.
The vertical line denotes the equilibrium geometry.
} 
\label{Plot_C2_NPE_VTZF12}
\end{figure}

In Fig.~\ref{Plot_C2_NPE_VTZF12} we plot the deviation between FCIQMC and FCIQMC-F12 energies 
with various basis sets and our reference FCIQMC-F12/cc-pVTZ values, as a function of bond length.
In Table~\ref{Tab:NPE} we present non-parallelity errors (NPE), defined as the maximum absolute deviations over the
range of geometries considered.
The F12 corrections reduce the NPE of the FCIQMC/cc-pVDZ method by more than a factor of seven, which is comparable with
the NPE of the FCIQMC/cc-pVQZ energies.
%The reduction in the NPE across this range is clear, with the F12 corrections reducing the NPE by a factor of over seven in the cc-pVDZ case, and resulting in a quality comparable, although just
%short of, the cc-pVQZ basis. 
However, the cc-pVQZ basis requires over 30$\times10^6$ walkers to converge the FCIQMC energy\cite{BoothC2}, even when when angular momentum symmetry is utilised. 
In contrast, the cc-pVDZ basis only requires a tenth of this number, even without the additional symmetries, highlighting the saving achieved with the hybrid F12 approach within FCIQMC.
The effect of including the whole range of bond lengths is unlikely to affect the NPEs, since the dominant basis-set incompleteness error is found at equilibrium
and compressed geometries, where the electrons are in close proximity, and dynamical correlation is at its largest.

%=======================================================
% C2 BINDING CURVES
%=======================================================
%\begin{figure}
%\begin{center}
%\includegraphics[scale=0.45]{Plots/Binding_Curves/plot_C2_VDZ_F12_BindingCurve}
%\end{center}
%\caption[The C$_2$ binding curves obtained with each basis set, compared to F12 corrected results]
%{A comparison of the C$_2$ binding curves calculated here.  The VDZ curve mostly describes the $X^1\Sigma_g^+$ state, 
%but relaxes into the lower energy $B^1\Delta_g$ state for $r = 1.7 - 2.4$ \AA, whereas the VTZ energies describe the 
%$X^1\Sigma_g^+$ state for all geometries except $r = 2.2$ \AA.  The VQZ binding curve was calculated by conserving 
%$M_l = 0$, and so entirely represents the diabatic $X^1\Sigma_g^+$ potential energy surface.  These energies were all 
%calculated from $\bra E_{\textrm{proj}} \ket_{\tau}$, or $\bra E_{\textrm{S}} \ket_{\tau}$ for $r \geq 1.6$.  
%Adding the $[2]_{\textrm{R12}}$ and $[2]_{\textrm{S}}$ corrections to the VDZ energies (VDZ+F12) results in a binding 
%curve very similar to VQZ.  Similar corrections to the VTZ values (VTZ+F12) provides even lower energies than the 
%VQZ and VDZ+F12 results.  The inset simply provides an enlargement of the equilibrium region to better discern the 
%energy differences.
%} 
%\label{Plot_C2_F12_BindingCurves}
%\end{figure}

%=======================================================
% NPE FROM VTZ+F12
%=======================================================

It is this fact which makes the calculation of the dissociation energy so sensitive to basis set incompleteness. The calculation was performed using the energy of the system at
a dissociated geometry of 6.21265 \AA\, rather than twice the atomic energy. This provided a highly multiconfigurational state with which to test the application of the corrections, although
the uncorrected FCIQMC energies were checked for any size-consistency error in this value relative to the atomic reference systems.
The results can be seen in Table~\ref{Tab:NPE}, compared to the experimental dissociation value which is approximately corrected for zero-point energy and core-valence correlation effects. A similar
trend to the NPEs is observed, with the F12 corrected FCIQMC value in a cc-pVDZ basis achieving accuracy just shy of the uncorrected cc-pVQZ result. 

\begin{table}
\begin{tabular}{l r@{}l r@{}l}
\hline
\multirow{2}{*}{Basis} & \multicolumn{2}{c}{NPE}                & \multicolumn{2}{c}{$D_{\textrm{e}}$} \\
                       & \multicolumn{2}{c}{mE$_{\textrm{h}}$}  & \multicolumn{2}{c}{/ kcal mol$^{-1}$} \\
\hline                                               
cc-pVDZ              &   86&.8(4)    &\hspace{12pt}130&.0(1)  \\ %&  1&.2725   &   1809  \\ % &   14.6   \\%   & 129.95(8)  \\
cc-pVTZ              &   21&.0(6)    &   139&.9(3)  \\ %&  1&.2525   &   1837  \\%    &   15.9   \\%   &    139.63(2)  \\
cc-pVQZ              &   4&.9(6)     &   143&.3(2)  \\ %&  1&.2466   &   1827  \\%    &   16.1   \\%   &    143.44(5)  \\
cc-pVDZ+F12          &   12&.3       &   142&.3  \\ %&  1&.2477   &   1853  \\%    &   14.9   \\%   &    142.35557  \\
cc-pVTZ+F12          &     &         &   145&.3  \\ %&  1&.2465   &   1851  \\%    &   16.4   \\%   &   145.037    \\
Expt.                &     &         &   146&.9(5) \\ % & 1&.2425    &   1855 \\%     &   13.3   \\%  &    146.9(5)      \\ 
\hline
\end{tabular}
\caption
{The non-parallelity error and dissociation energies obtained from each of the C$_2$ binding curves.  
The uncorrected cc-pVQZ results were taken from Ref.~\onlinecite{BoothC2}.  
The NPE refers to 
the non-parallelity error across the $X^1\Sigma_g^{+}$ state up to geometries of 1.6\AA.
%The dissociation energies ($D_{\textrm{e}}$), equilibrium bond lengths ($r_{\textrm{e}}$), and 
%harmonic vibrational frequencies ($\omega_{\textrm{e}}$) were calculated by fitting the $0.9 \leq r \leq 1.6$ region 
%of each curve to an even-tempered Gaussian expansion \cite{BytautasFit}.  
The experimental dissociation energy was taken from Ref.~\onlinecite{RuedenbergCEEIS}, and includes
a correction to account for core-valence correlation.
Random errors are not included in the F12 corrected values, but can be expected to have errors of at least the errors in the uncorrected energies.
}
%and the experimental $r_{\textrm{e}}$ and $\omega_{\textrm{e}}$ from Huber and Herzberg \cite{HuberHerz}.} 
\label{Tab:NPE}
\end{table}

Despite the large improvement of the cc-pVTZ basis FCIQMC-F12 result, the dissociation energy is still not within chemical accuracy of the experimental value.  
When taking into account random errors of the FCIQMC method, as well as uncertainty in experimental results, these values differ by 1.6(6)kcalmol$^{-1}$.
%Despite the large improvement of the cc-pVTZ basis FCIQMC-F12 result, the result is still not inside chemical accuracy of the experimental value, which when taking into account random errors of the FCIQMC
%method, as well as uncertainty in experimental results, are in error by 1.6(6)kcalmol$^{-1}$. 
From a calculation of the all-electron cc-pVTZ FCIQMC-F12 energy at the experimental equilibrium and
fully dissociated geometries, the error is reduced to 0.92(53)kcalmol$^{-1}$, with the approximate core-valence correction neglected. Although this improves the final dissociation energy to within
`chemical accuracy', it is yet to be seen whether remaining basis set error (such as a need for additional diffuse functions within the F12 framework), other neglected effects, or experimental
ambiguity can account for the remaining discrepancy between the two values. However, it is clear from this, and other studies\cite{RuedenbergCEEIS}, that achieving chemical accuracy for this system is
exceedingly difficult, where strong correlation, as well as significant basis set incompleteness can cause severe errors.

\section{Summary and Conclusions}

In this paper we use an {\em a posteriori} $\tworonetwo$ and $\twos$ approach\cite{Torheyden:JCP131-171103,Kong:JCP135-214105,Kong:JCP133-174126} in order to account for 
basis set incompleteness within FCIQMC. This allows for perturbative 
excitations from the sampled FCIQMC wavefunction into a space of strongly orthogonal geminal functions, via the two-body reduced density matrix which is accumulated on-the-fly. 
The basis set error in the one-particle space is accounted for via an adaptation of the $\twos$ method designed for incompleteness in CASSCF. 
It is shown that the need for an F12 correction which can correlate a multireference zeroth order wavefunction, is crucial in strongly interacting systems. 
The convergence of the F12 properties with respect
to both increasing walker number, and sampling time of the density matrix is investigated, and found to converge at a similar rate to that of the projected energy estimate used in $i$-FCIQMC.  This was the case 
for both the primarily dynamically correlated carbon dimer at equilibrium geometry, as well as the highly multiconfigurational stretched geometry. A relatively modest increase
in computational effort was required to calculate these F12 corrections, and therefore a large saving was achieved by using the corrections to reduce the basis set error with FCIQMC in converged calculations.

This machinery was then used to treat the carbon dimer at a variety of bond lengths and basis sets, in order to quantify the benefit of the F12 corrections. This system was the
subject of a previous investigation, where remaining basis set error, even in quadruple zeta basis sets, prevented convergence to within chemical accuracy. In keeping with other studies,
the F12 corrections were found to provide a gain of close to two cardinal numbers in the basis set when considering non-parallelity errors and dissociation energies, and 
provided a balanced description of both dynamic and strong correlation effects at equilibrium and stretched geometries. This enabled the calculation of the dissociation energy
to within chemical accuracy once core electron correlation was explicitly taken into account. In the future, we hope this approach will allow us to tackle a range of strongly correlated
systems, without requiring large computational basis sets for the FCIQMC calculation. This approach however does not allow for a relaxation of the FCIQMC wavefunction due to the presence of
the geminal functions. A future aim will be to compare this approach with an alternative {\em a priori} transcorrelation of the Hamiltonian\cite{Yanai:JCP136-084107,Handy69}, 
as well as more traditional CASPT2 and MRCI techniques.

\section*{Acknowledgements}
The authors wish to thank Andreas K{\" o}hn for the use of his integral transformation code, and Ed Valeev for providing comparison numbers and useful discussions, as well as acknowledging financial
support from the EPSRC (through grant numbers EP/I014624/1 and EP/J003867/1), Trinitiy College, Cambridge (GHB), the Woolf--Fisher Trust, New Zealand (DMC), and the Royal Society (DPT).

%\bibliography{F12Bib}

%

\end{document}